\documentclass
[prl,twocolumn,10pt,tightenlines,showpacs,letterpaper,runinaddress]{revtex4}
\usepackage{graphicx}
\begin{document}
\title[Universal charge-radius relation]{Universal charge-radius relation for subatomic and astrophysical compact objects}
\author{Jes Madsen}
\affiliation{Department of Physics and Astronomy, University of Aarhus, DK-8000 \AA rhus C, Denmark}

\begin{abstract}
Electron-positron pair creation in supercritical electric fields limits the
net charge of any static, spherical object, such as superheavy nuclei,
strangelets, and Q-balls, or compact stars like neutron stars, quark stars,
and black holes. For radii between $4\times10^{2}$ fm and $10^{4}$ fm the
upper bound on the net charge is given by the universal relation
$Z=0.71R_{\text{fm}}$, and for larger radii (measured in fm or km)
$Z=7\times10^{-5}R_{\text{fm}}^{2}=7\times10^{31}R_{\text{km}}^{2}$. For
objects with nuclear density the relation corresponds to $Z\approx0.7A^{1/3}$
($10^{8}<A<10^{12}$) and $Z\approx7\times10^{-5}A^{2/3}$ ($A>10^{12}$), where
$A$ is the baryon number. For some systems this universal upper bound improves
existing charge limits in the literature.

\end{abstract}
\date{September 17, 2007; to appear in PRL April 2008}
\maketitle

Spontaneous formation of real electron-positron pairs in supercritical
electric fields is known to lead to screening of highly charged objects
\cite{mr75,gmr85}. This is in addition to the vacuum polarization effect 
caused by virtual pairs \cite{gmr85}. In the following it will be shown, 
that pair formation implies
a universal upper limit on the net charge of any static, spherical object of
given radius. The upper bound on the net charge for $R>4\times10^{2}$ fm is
$Z_{\infty}=0.71R_{\text{fm}}=7.1\times10^{17}R_{\text{km}}$, where the radius
is measured in fm or km to show the typical scales for subatomic objects or
compact stars like neutron stars, quark stars and black holes, and the
subscript $\infty$ indicates infinite time available for pair formation.
Hypothetical superheavy nuclei, quark nuggets (also known as strangelets), and
other objects with nuclear matter density have $R_{\text{fm}}\approx A^{1/3}$,
where $A$ is the baryon number, so the charge bound can be recast as
$Z_{\infty}\approx0.7A^{1/3}$. This universal charge bound is complementary to
and often significantly more restrictive than other bounds in the literature.
However, it implicitly assumes that infinite time is available to populate
electron levels via pair creation. Taking the relevant timescales into account
the maximum charge increases for $R>10^{4}$ fm ($A>10^{12}$) to $Z\approx
7\times10^{-5}R_{\text{fm}}^{2}\approx7\times10^{-5}A^{2/3}$ (the latter
expression assuming nuclear matter density). This relation improves existing
limits for e.g. strangelets, reproduces an earlier result for massive black
holes, but is inferior to charge limits based on stellar stability for
gravitationally bound neutron stars when only finite time is available for
pair creation.

In the following the upper bound on the charge of spherical objects will be
derived in the context of a relativistic Thomas-Fermi model calculation
following the approach of M\"{u}ller and Rafelski \cite{mr75}. The universal
relation will be analytically derived in the high mass, infinite time limit,
and as typical examples homogeneously charged superheavy nuclei and
color-flavor locked strangelets dominated by surface charge will be shown
numerically to approach the universal charge-radius relation. Finite time for
pair creation will then be taken into account and the corresponding (higher)
charge limits derived. The results will be compared to earlier charge-mass
relations for strangelets, and to limits for astrophysical objects. Whereas
the derivations do not include general relativity it is argued, that the
universal charge-radius limit should also apply approximatively to black
holes, and in fact the results agree with a limit derived for black holes in
general relativity.

In a continuous approximation at zero temperature (more about finite
temperature effects later) the number density of electrons, $n_{\text{e}}$,
is given by the electron mass, $m_{\text{e}}$, and Fermi energy,
$\mu_{\text{e}}$, via

\begin{equation}
n_{\text{e}}=\frac{\left[  \left(  \mu_{\text{e}}^{\text{eff}}+e\phi\right)
^{2}-m_{\text{e}}^{2}\right]  ^{3/2}}{3\pi^{2}}\theta\left(  \mu_{\text{e}
}^{\text{eff}}+e\phi-m_{\text{e}}\right)  .
\end{equation}

Here the effective chemical potential $\mu_{\text{e}}^{\text{eff}}
=\mu_{\text{e}}-e\phi$, where $\phi$ is the electric potential, and the
$\theta$-function takes into account that $\mu_{\text{e}}$ must exceed the
electron mass. In the usual Thomas-Fermi model describing neutral atomic
systems one takes the energy needed to add an additional electron,
$\mu_{\text{e}}^{\text{eff}}=m_{\text{e}}$. In the present context the focus
is on maximally charged rather than neutral systems, and here one instead
takes $\mu_{\text{e}}^{\text{eff}}=-m_{\text{e}}$, corresponding to the top of
the negative energy sea \cite{mr75}. This represents a situation where all
levels accessible to spontaneous vacuum decay are filled, and as shown in
\cite{mr75,gmr85} it reproduces the results of single-particle calculations carried
out for core charges up to a few hundred that demonstrate how more and more
real electron states dive into the negative energy continuum as the core
charge increases, leading to the creation of a negatively charged vacuum. Thus
\begin{equation}
n_{\text{e}}=\frac{\left[  \left(  e\phi\right)  ^{2}-2m_{\text{e}}
e\phi\right]  ^{3/2}}{3\pi^{2}}\theta\left(  e\phi-2m_{\text{e}}\right)  .
\end{equation}

In regions of space containing only electrons Poisson's equation is given by
$\nabla^{2}(e\phi)=4\pi e^{2}n_{\text{e}}$, or in dimensionless units, where
the radial coordinate $\xi\equiv m_{\text{e}}r$, and the normalized potential
$y\equiv e\phi/m_{\text{e}}$, with $\alpha\equiv e^{2}$,
\begin{equation}
\nabla^{2}y=\frac{4\alpha}{3\pi}\left[  y^{2}-2y\right]  ^{3/2}\theta(y-2).
\end{equation}
In regions with additional charge, a corresponding source term must be added
to the right-hand side of Poisson's equation. Two illustrative types of core
charge distributions have been considered: A uniformly charged spherical core
(like a nucleus) and a uniformly charged spherical shell with no internal
charge (like a color-flavor locked quark matter lump or a large, perfectly
conducting nuclear matter system, where the net core charge moves to the
surface). Together these idealized distributions span the likely range of real
distributions, and as shown in the following, they lead to identical results
for the screened charge in the limit of large system radius.

Boundary conditions for Poisson's equation are $\nabla y\rightarrow0$ for
$\xi\rightarrow0$ and $y\rightarrow0$ for $\xi\rightarrow\infty$. But the
actual behavior of $y$ for large $\xi$ is explicitly known in the maximally
charged case. Denote the total (positive) core charge by $Z_{\text{core}}$ and
the total number of electrons by $N_{\text{e}}$. Then $Z_{\infty
}=Z_{\text{core}}-N_{\text{e}}>0$ is the net charge of the maximally charged
system seen by an observer outside the radius $\xi_{2}$, where the electron
density drops to zero, $y(\xi_{2})=2$ (corresponding to $e\phi=2m_{\text{e}}
$). But to this observer
\begin{equation}
y=\frac{Z_{\infty}\alpha}{\xi}\text{ for }\xi\geq\xi_{2}\text{.}
\end{equation}

Using this relation for $\xi=\xi_{2}$ gives
\begin{equation}
Z_{\infty}=\frac{2m_{\text{e}}r_{2}}{\alpha},
\end{equation}
where $r_{2}$ denotes the physical radius corresponding to $\xi=\xi_{2}$. In
the bulk of a homogeneously charged core electrons neutralize the local core
charge, and a deviation from local charge neutrality occurs only very close to
the surface. As confirmed by numerical solutions of Poisson's equation the
characteristic width of the screening electron cloud outside the core charge
is of order $m_{\text{e}}^{-1}$ or a few hundred fm (deviations from local
charge neutrality occurs within a similar zone inside the surface). Thus in
the limit of $r_{2}\gg m_{\text{e}}^{-1}$ the width is small compared to the
physical radius of the core, $R$, and to good approximation the net charge of
a maximally charged spherical system with $R\gg m_{\text{e}}^{-1}
\approx4\times10^{2}$ fm is therefore (since $R\approx r_{2}$)
\begin{equation}
Z_{\infty}\approx\frac{2m_{\text{e}}R}{\alpha}=0.71R_{\text{fm}}
=7.1\times10^{17}R_{\text{km}}\text{.}
\end{equation}
For objects with density of order nuclear matter density $R_{\text{fm}}\approx
A^{1/3}$. Therefore, for such objects charge and baryon number are related by
\begin{equation}
Z_{\infty}\approx0.7A^{1/3}\text{ for }A\gg10^{8}\text{.}
\end{equation}

Figure 1 shows how well the universal charge-radius relation fits numerical
solutions to the relativistic Thomas-Fermi model at large radius. The example
here is for color-flavor locked strangelets composed of a core of equal
numbers of up, down, and strange quarks, leading to zero quark charge density
in the bulk of the core, but with a core surface charge due to surface
depletion of strange quarks, so that $Z_{\text{core}}\approx0.3A^{2/3}$
\cite{madsen2001} and radius $R_{\text{fm}}\approx1.1A^{1/3}$ (such a system
has a discontinuity in $dy/d\xi$ at the core surface in addition to the
boundary conditions previously described for the Poisson equation). One
clearly sees how the actual solution of Poisson's equation for the net
screened charge shifts from $Z_{\infty}\approx Z_{\text{core}}$ to the
universal relation $Z_{\infty}\approx2\xi/\alpha$ when going from small
($\xi\ll1$) to large ($\xi\gg1$) radii. Similar behavior is found for systems
with other $Z_{\text{core}}(A)$ dependence and regardless of whether the core
charge is uniformly distributed, or distributed as a surface charge as in the
case of strangelets and for cores that behave as ideal conductors. In all
cases tested, the maximal charge approaches the \textit{same} universal
charge-radius relation $Z_{\infty}\approx0.71R_{\text{fm}}$ for $m_{\text{e}
}R\equiv\xi\gg1$ (physical radius $R$ exceeding $4\times10^{2}\,$fm) as expected.

So far the calculations have assumed a static situation with infinite time
available to fill all accessible electron states. However the pair formation
process involves tunneling and the rate for this (number of pairs produced per
volume per time) has been shown to be \cite{Schwinger}
\begin{equation}
W=\frac{m_{\text{e}}^{4}}{4\pi^{3}}\left(  \frac{E}{E_{\text{cr}}}\right)
^{2}\sum\limits_{n=1}^{\infty}\frac{1}{n^{2}}\exp\left(  -n\pi\frac
{E_{\text{cr}}}{E}\right)  ,
\end{equation}
where the critical field is $E_{\text{cr}}=m_{\text{e}}^{2}/\alpha
^{1/2}\approx1.3\times10^{16}$V/cm. Taking $E=Z\alpha^{1/2}/R^{2}$ to
represent the \textquotedblleft surface region\textquotedblright\ of the
spherical charge distribution studied here, and including only the leading
$n=1$ term in the sum, this corresponds to
\begin{equation}
W\approx\frac{m_{\text{e}}^{4}\alpha^{2}Z^{2}}{4\pi^{3}\xi^{4}}\exp\left(
-\pi\frac{\xi^{2}}{\alpha Z}\right)  .
\end{equation}

Taking the active charge producing layer to have a volume $V=4\pi R^{2}\Delta
R$, with thickness $\Delta R\approx m_{\text{e}}^{-1}$, assuming $WV=-dZ/dt$,
and defining a characteristic timescale $\tau$ for charge equilibration from
$dZ/Z\equiv-dt/\tau$, this leads to a timescale
\begin{equation}
\tau=\frac{\pi}{\alpha m_{\text{e}}}x\exp(x)=5.5\times10^{-19}\text{s}
\,x\exp(x),
\end{equation}
where $x\equiv\pi\xi^{2}/\alpha Z$. Because of the exponential a finite time
available for pair production corresponds to a (roughly) fixed value of $x$
for a wide range of $\tau$ (it also means that the results do not depend
crucially on the assumptions about $V$; taking $V=4\pi R^{3}/3$ instead of the
surface layer chosen makes little difference to the charge relations below).
In other words, pair production in a finite time gives a maximum charge value
which is proportional to $\xi^{2}$, rather than $\xi$ as were the case with
infinite time available. For example, for $\tau=1$ s the maximally allowed
charge becomes
\begin{equation}
Z_{1\text{s}}\approx11.2\xi^{2}\approx7\times10^{-5}R_{\text{fm}}^{2}
\approx7\times10^{-5}A^{2/3}\text{,}
\end{equation}
where the last equality assumes nuclear matter density. For $\tau=10^{10}$
years one gets very similar results, except for dividing the numerical
prefactors by $2.0$, whereas a typical weak interaction timescale of
$10^{-10}$ s corresponds to multiplication of the numerical prefactors by
$2.4$. Concentrating on $Z_{1\text{s}}$ we therefore have two regimes for the
universal charge-radius relation. For $Z_{1\text{s}}(\xi)<Z_{\infty}(\xi)$
pair production is sufficiently rapid to fill the electron states as assumed
in the relativistic Thomas-Fermi model, and therefore $Z\approx Z_{\infty
}\approx2\xi/\alpha$ for $1<\xi<25$. For $Z_{1\text{s}}(\xi)>Z_{\infty}(\xi)$
($\xi>25$) we have instead $Z\approx Z_{1\text{s}}\approx11.2\xi^{2}$. The
corresponding division between the two regimes in terms of radius and baryon
number (assuming nuclear matter density) is $R_{\text{fm}}\simeq
1.0\times10^{4}$ and $A\approx10^{12}$ respectively. The higher charge
permitted for macroscopic spheres because of the finite time effect explains
why a Van de Graaff generator can work at potential in excess of $1$ MV as
would be the limit with infinite time to screen the charge with electrons from
pair production.

The existence of a universal maximal charge-radius relation for spherical
objects regardless of their physical nature is interesting in its own right,
but it also has applications to several areas of subatomic physics and
astrophysics as outlined in the following.

It has been speculated that there could exist branches of metastable
superheavy nuclei. If so, the charge of the maximally ionized superheavy ion
should obey the relation $Z\approx0.7A^{1/3}$ ($10^{8}<A<10^{12}$) or
$Z\approx7\times10^{-5}A^{2/3}$ ($A>10^{12}$). Because the nuclear charge
increases almost linearly with $A$, charge screening becomes important already
for $A\approx10^{3}$ (as also pointed out in \cite{mr75}), but numerical
solutions of Poisson's equation show that one needs $\xi>1$ or $A>10^{8}$ for
the universal relation to become quantitatively accurate.

The possible existence of metastable or even stable quark nuggets or
strangelets has been widely discussed \cite{bodmer71}. For large chunks of
quark matter, which could exist in our Galaxy as a result of binary compact
star collisions, the $(Z,A)$-relation in the high-mass limit has so far been
taken as either $Z\approx8A^{1/3}$ (for non-color-flavor locked strangelets
with $A\lesssim10^{7}$ \cite{heiselberg}), or $Z\approx0.3A^{2/3}$ (for
color-flavor locked strangelets regardless of $A$ \cite{madsen2001}). The
possible importance of charge screening due to supercritical field electrons
was mentioned by Farhi and Jaffe \cite{farhi84}, and a first numerical study
was included in \cite{madsenlarsen2003}. From the discussion above it follows
that the universal relation $Z\approx0.7A^{1/3}$ ($10^{8}<A<10^{12}$) and
$Z\approx7\times10^{-5}A^{2/3}$ ($A>10^{12}$) can be applied for $A>10^{8}$ as
an upper envelope on the strangelet charge regardless of the details of the
quark phase, and for color-flavor locked quark matter this envelope represents
the actual maximum ionization state possible, since the \textquotedblleft
envelope\textquotedblright\ value of $Z$ is smaller than the core charge
$0.3A^{2/3}$.

Q-balls have been suggested in various varieties, some of which can be charged
\cite{Q-balls}. Again, the universal relation derived here should apply.

Turning to astrophysical objects, it is normally assumed that compact stars
such as neutron stars and quark stars are close to electrical neutrality,
since a net positive charge would attract electrons from the interstellar
medium. However, many papers have dealt with the theoretical possibility of
non-zero charge and its influence on stellar structure, and limits (typically
from arguments related to stellar stability) have been placed on the charge
allowed. From a comparison of gravitational binding and electric repulsion,
several authors (e.g. \cite{glendenning}) have found that the net charge of
gravitationally bound stars is limited by $Z<10^{-36}A$, where the baryon
number $A\approx10^{57}$ for a typical compact star, thus $Z<10^{21}$. From
the universal relation derived above, we have the limit $Z_{1\text{s}}
<7\times10^{-5}A^{2/3}$, which is conceptually very different, but happens to
give a larger number for $A\approx10^{57}$, namely $Z_{1\text{s}}
<7\times10^{33}$. This shows that stabilization of charged neutron stars
should result from trapping of real electrons from the surroundings; there is
not enough time to ensure sufficient numbers of electrons from pair creation.
With infinite time available, the static maximal charge would be $Z_{\infty
}\approx0.7A^{1/3}\approx10^{19}$, i.e. stable with respect to the electric
repulsion, but to approach this limit requires orders of magnitude more time
than the age of the Universe if one were to start out with a much higher charge.

Strange stars consisting of quark matter are self-bound due to strong
interactions in addition to gravity, so the relevant stability limit here
comes from a comparison of the Coulomb energy and the total (strong
interaction) binding energy per baryon. Stability requires $Z^{2}\alpha/AR$ to
be less than a few MeV, which is easily satisfied (by 8 orders of magnitude)
for $Z=Z_{1\text{s}}$. Therefore the charge of strange stars may in principle
saturate the universal limit $Z_{1\text{s}}\approx7\times10^{-5}A^{2/3}$, and
such a system could have interesting properties related to pair creation even
at zero temperature in addition to the (much faster) finite temperature pair
creation phenomena discussed in \cite{Usov}.

Black holes are formally exempt from the treatment above since they are
described by general relativity. However, formation of an astrophysical black
hole involves the collapse of a mass (and for the present purpose charge) that
seen from an outside observer only asymptotically reaches the horizon.
Therefore one might expect that the universal charge-radius relation would
indeed be obeyed by the black hole because it would be obeyed during the
formation process, and the maximal net charge seen from the outside again
should not exceed $7\times10^{31}R_{\text{km}}^{2}$. Studies of pair formation
in a general relativistic description of black holes has in fact led to
results almost identical to this \cite{ruffini}, a result which is many orders
of magnitude below other limits derived for the maximal charge of a stable
black hole as long as the mass is below $10^{8}$ solar masses \cite{ruffini}.

The derivations above were all based on zero temperature relations.
These relations remain valid for temperature $T\ll \mu_{\text e}-m_{\text e}
=e\phi -2m_{\text e}$, so at low temperatures thermal electrons can be
neglected except near the edge of the zero temperature maximally charged
configuration, where $e\phi \rightarrow 2m_{\text e}$. 
The lowest order thermal contribution to the electron number
density in this regime is roughly $m_{\text e}^{3/2}T^{3/2}$, giving a 
thermal contribution to the total charge which is suppressed by
$(T/m_{\text e})^{3/2}$ relative to the zero temperature charge
$Z_{1\text{s}}$.
Therefore the zero temperature relations provide good approximations
for $T\ll m_{\text e}$.

Another caveat in applying the universal charge-radius relation derived above to
real subatomic or astrophysical objects is related to the tunneling timescale
involved. The larger the system, the lower is the probability of filling all
electron levels on a short timescale. Therefore, even though the universal
relation is independent of the internal composition and structure of the
spherical core, the realization of the universal behavior in realistic systems
will depend somewhat on the nature of the object and its previous history.
Systems with a high electrical conductivity (e.g. nuclear matter or quark
matter) will rearrange any initial net charge in a way so that the charge gets
concentrated in a thin surface layer. The small width of this layer and the
surrounding electron atmosphere improves chances for the universal
charge-radius relation to be realized. Other systems, such as strangelets or
quark stars composed of color-flavor locked quark matter, are electrically
neutral in the bulk quark phase already \cite{rajagopal}, and have only a net
quark charge on the surface \cite{madsen2001,usov2004}, 
so here the electron shield is thin almost
\textquotedblleft by construction\textquotedblright. Positively charged
objects in general tend to neutralize by trapping electrons from the
surroundings, e.g. from the interstellar medium. Therefore, most of the bulk
of macroscopic objects will be close to charge neutral, and the interesting
physics will be related to processes ionizing such objects. Such ionization
will take place \textquotedblleft from the outside\textquotedblright\ and
involve a restricted radial range, again minimizing the timescale problem.

In conclusion, it has been demonstrated that electrons created in
supercritical fields lead to a universal relation between the maximal charge
and radius of any static, spherically symmetric object with a size exceeding a
few hundred fm. The relation limits the charges of objects such as superheavy
nuclei, strangelets, Q-balls, neutron stars, quark stars, and black holes. For
some of the objects studied, the new limits are more restrictive than other
charge relations and limits existing in the literature.

This work was supported by the Danish Natural Science Research Council.

\begin{figure}
[b]
\begin{center}
\includegraphics[
angle=-90.,
width=3.4in
]%
{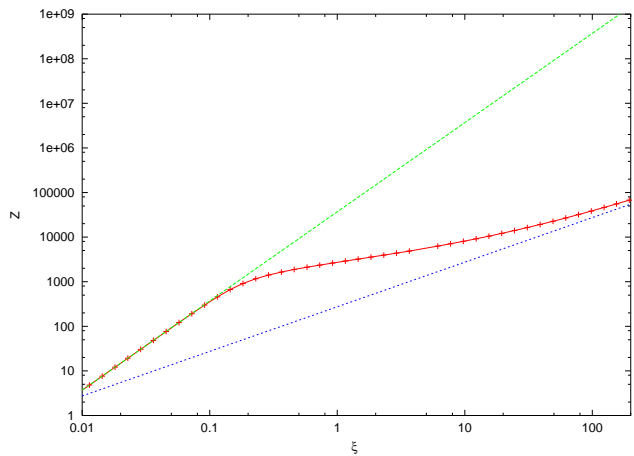}%
\caption{Charge as a function of scaled radius, $\xi$. The lower (blue,
dotted) line represents the universal charge-radius relation $Z_{\infty
}\approx2\xi/\alpha$ valid for large radii and infinite time. The upper
(green, dashed) line is the quark charge of color-flavor locked strangelets,
and the middle points with fitted (red) curve shows the maximal net charge,
$Z$, as calculated numerically from the relativistic Thomas-Fermi model. As
can be clearly seen, the electron screening due to the supercritical electric
field becomes important for $\xi>0.1$, and the universal charge-radius
relation is a good representation for $\xi\gg1$.}
\end{center}
\end{figure}
\end{document}